\documentclass[aip,apl,preprint,groupedaddress]{revtex4-1}
\usepackage[dvips]{graphicx}
\begin{document}

\title{Systematic study on transport properties of FeSe thin films with various degrees of strain}

\author{Fuyuki Nabeshima}
\email[]{nabeshima@maeda1.c.u-tokyo.ac.jp}
\affiliation{Department of Basic Science, the University of Tokyo, Meguro, Tokyo 153-8902, Japan}

\author{Masataka Kawai}
\affiliation{Department of Basic Science, the University of Tokyo, Meguro, Tokyo 153-8902, Japan}

\author{Tomoya Ishikawa}
\affiliation{Department of Basic Science, the University of Tokyo, Meguro, Tokyo 153-8902, Japan}

\author{Atsutaka Maeda}
\affiliation{Department of Basic Science, the University of Tokyo, Meguro, Tokyo 153-8902, Japan}

\date{\today}

\begin{abstract}
We performed systematic studies on the transport properties of FeSe thin films with controlled degrees of in-plane lattice strain, including both tensile and compressive strains.
The superconducting transition temperature, $T_{\mathrm c}$, increases up to 12 K for films with compressive strain while the superconductivity disappears for films with large tensile strains. 
On the other hand, the structural (nematic) transition temperature, $T_{\mathrm s}$, slightly decreases as the in-plane strain is more compressive. 
This suggests that the structural transition can be extinguished by a smaller amount of Te substitution for films with more compressive strain, which may lead to higher $T_{\mathrm c}$ in FeSe$_{1-x}$Te$_x$.
It was also found that the carrier densities evaluated via transport properties increase as the in-plane strain becomes more compressive. 
A clear correlation between $T_{\mathrm c}$ and the carrier densities suggests that it is essential to increase carrier densities for the $T_{\mathrm c}$ enhancement of iron chalcogenides.

\end{abstract}

\pacs{}
%\keywords{}
\maketitle

Iron-based superconductors (FeBSs) are attracting widespread interest in the field of superconductivity in terms of both fundamental and applied research.
An iron chalcogenide superconductor, FeSe$_{1-x}$Te$_x$,\cite{Wu08,Fang08} is a FeBS with the simplest crystal structure, consisting of conducting planes alone.
The parent material FeSe shows a structural transition from a tetragonal to an orthorhombic phase at 90 K,\cite{PhysRevLett.103.057002} while it exhibits no magnetic order at ambient pressure, different from many other parent materials of FeBSs.
The structural transition has an electronic origin,\cite{PhysRevB.90.121111,PhysRevLett.113.237001} and is also called the nematic transition.
FeSe exhibits superconductivity below 9 K.
The Te substitution of Se increases $T_{\mathrm c}$; the highest $T_{\mathrm c}$ of 14 K is obtained at $x=0.5$ at ambient pressure.\cite{Fang08}
In spite of its low $T_{\mathrm c}$ compared with other iron-based superconductors, it maintains large critical current densities, $J_{\mathrm c}$, at very high magnetic fields,\cite{Tsukada11,NCommun.4.1347,NatCommun.7.13036} and thus, FeSe$_{1-x}$Te$_x$ has great potential in high-field applications such as superconducting wires and tapes for magnets.

Much research on the thin-film growth of FeSe$_{1-x}$Te$_x$ has focused on the composition of $x=0.5$---the optimal composition for achieving high $T_{\mathrm c}$ in bulk samples---and enhanced superconductivity due to the in-plane lattice strain ($T_{\mathrm c}^{\mathrm {onset}}$ $\approx$ 19 K at the maximum\cite{Bellingeri10}) has been reported.\cite{Bellingeri10,Tsukada11,APL.99.202503,NatCommun.4.1347}
However, the optimal composition of film samples is different from that of bulk samples.
It is well-known that bulk samples with $0.1<x<0.4$ of FeSe$_{1-x}$Te$_x$ are not available because of phase separation. 
We have previously demonstrated that single crystalline films of FeSe$_{1-x}$Te$_x$ in the whole composition region, including the bulk phase separation region, can be grown via pulsed laser deposition and that the true optimal composition of film samples is in the phase separation region of $0.1<x<0.4$.\cite{yi15pnas,yiSciRep17}

We have grown FeSe$_{1-x}$Te$_x$ films on two different substrates, namely, CaF$_2$ and LaAlO$_3$ (LAO), and have obtained two electronic phase diagrams for these samples.
Although, the detailed behaviors of the two phase diagrams are different from each other, they show the same behavior qualitatively: 
(i) Te substitution decreases the nematic (structural) transition temperature, $T_{\mathrm s}$, 
(ii) when the nametic transition disappears, $T_{\mathrm c}$ rapidly increases, and the highest $T_{\mathrm c}$ is obtained just after the disappearance of the nematic transition, 
and (iii) $T_{\mathrm c}$ decreases monotonically with increasing $x$ in the tetragonal phase.
The compositions, $x_{\mathrm c}$, where the structural transition disappears are different between the two phase diagrams; $x_{\mathrm c}$ is smaller for films on CaF$_2$ ($x_{\mathrm c}=0.2$ for CaF$_2$ and $x_{\mathrm c}=0.4$ for LAO).

The fact that the sudden increase in $T_{\mathrm c}$ at $x_{\mathrm c}$ is commonly observed in the two different phase diagrams strongly indicate that the nematicity suppresses the superconductivity in FeSe$_{1-x}$Te$_x$.
Because the maximum value of $T_{\mathrm c}$ is higher for films on CaF$_2$ than on LAO, we expect the maximum value of $T_{\mathrm c}$ to become larger if we can suppress the nematicity faster.
Therefore, it is essential to elucidate the origin of the difference in $x_{\mathrm c}$ between films on CaF$_2$ and on LAO for further enhancement of superconductivity in FeSe$_{1-x}$Te$_x$.
From a simplistic point of view, the origin of the difference in the phase diagrams between films on the two substrates is expected to be in the difference in the strength of the lattice strain.
Indeed, there is a tendency that the $a$-axis length of films on CaF$_2$ is shorter than that on LAO.
Therefore, it is important to investigate the effects of strain on the physical properties of these materials.

In this letter, we report on the systematic studies of transport properties of FeSe thin films with various degrees of strain.
We demonstrate that the structural transition is suppressed by compressive strain, consistent with the fact that the suppression of $T_{\mathrm s}$ is stronger in films on CaF$_2$ than on LAO.
This result suggests that stronger compressive strain can make $x_{\mathrm c}$ smaller, which may result in further enhancement of $T_{\mathrm c}$ in FeSe$_{1-x}$Te$_x$ thin films.
In addition, we report a clear correlation between $T_{\mathrm c}$ and the carrier concentrations, which suggests that it is essential to increase carrier densities for realizing high $T_{\mathrm c}$ in FeSe.
Our results will provide important clues for further enhancement of $T_{\mathrm c}$ as well as the superconducting mechanism in iron chalcogenides.

All the FeSe thin films were grown by a pulsed laser deposition method using a KrF laser.\cite{Imai09,Imai10}
To change the strength of in-plane strain we grew films on three different substrates in this study, namely, LaAlO$_3$ (LAO), (LaAl)$_{0.7}$-(SrAl$_{0.5}$Ta$_{0.5}$)$_{0.3}$O$_3$ (LSAT), and LSAT with a LAO buffer layer.
The purpose of the use of the LAO buffer layers on LSAT is to avoid possible diffusion of oxygen at the interface of film and LSAT substrates.\cite{hanawa12}
However, it turned out that there were no significant differences in the crystalline quality and the superconducting properties between films on LSAT with and without LAO buffer layer.
The crystal structures and the orientations of the films were characterized with a four-circle X-ray diffractometer with Cu K$\alpha$ radiation at room temperature.
The thicknesses of the samples were evaluated with a Dektak 6 M stylus profiler. 
The electrical resistivity and the Hall resistivity were measured using a physical property measurement system from 2 to 300 K under magnetic fields up to 9 T.

\begin{figure}
\includegraphics[width=\linewidth, bb=0 0 430 398]{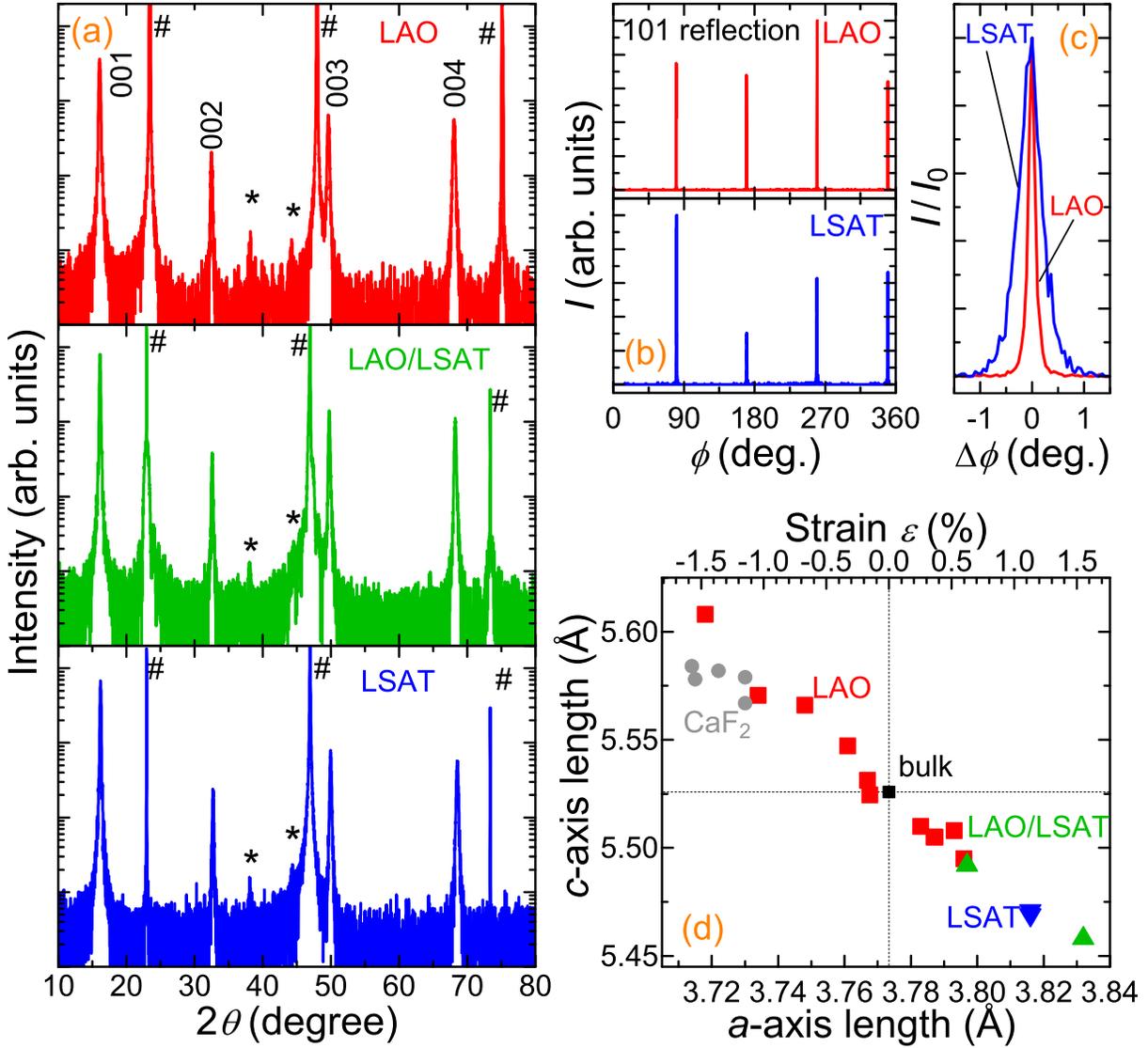}%
\caption{(a) XRD patterns of typical three FeSe films on different substrates: LAO, LSAT with LAO buffer, and LSAT. The number and asterisk signs represent peaks from the substrates and the Ag paste, respectively. Peaks of the LAO buffer layer overlap with those of LSAT. (b) $\phi$-scans of the 101 reflections of FeSe films on LAO (top) and LSAT (bottom). (c) Comparison of the 101 reflections of FeSe films on LAO and LSAT. (d) Relation between $a$- and $c$-axis length of the grown films. Data of bulk samples\cite{PRB.79.014522} and films on CaF$_2$ substrates\cite{Nabe13,Maeda14} are also plotted.}
\label{G-XRD}
\end{figure}

Figure \ref{G-XRD}(a) shows the X-ray diffraction patterns of three typical samples on different types of substrates.
Except for peaks from the substrates, buffer layers, and Ag paste for the resistivity measurements, all the peaks are identified as $00l$ reflections of FeSe, indicating the $c$-axis orientations of the films.
In-plane orientations of the films were also confirmed by the $\phi$-scans of the 101 reflections (Fig. \ref{G-XRD}(b)), which showed clear four-fold symmetry patterns.
Note that films on LSAT showed broader peak width of the 101 reflection than film on LAO; FWHM of the peak is $\sim 0.15^\circ$ for films on LAO and $\sim 0.5^\circ$ for films on LSAT.
The difference in the peak width may be due to the difference in the lattice mismatch between film and substrate.
We summarize the lattice constants of the grown films in Fig. \ref{G-XRD}(d). 
A clear negative correlation was observed between the $a$- and $c$-axis lengths, including those of bulk crystals,\cite{PRB.79.014522} which can be explained by the Poisson effect in crystals under in-plane strain.
These results demonstrate the successful growth of single crystalline FeSe films with varied degrees of strain in a wide range.
In terms of the strain parameter, $\varepsilon \equiv (a_{\mathrm {film}}-a_{\mathrm {bulk}})/a_{\mathrm {bulk}}$, we were able to obtain samples with $ -1.5\% < \varepsilon < 1.5\%$ in this study. 
%(CaF2膜の格子定数が小さいことに言及する？)

\begin{figure}
\includegraphics[width=\linewidth, bb=0 0 430 268]{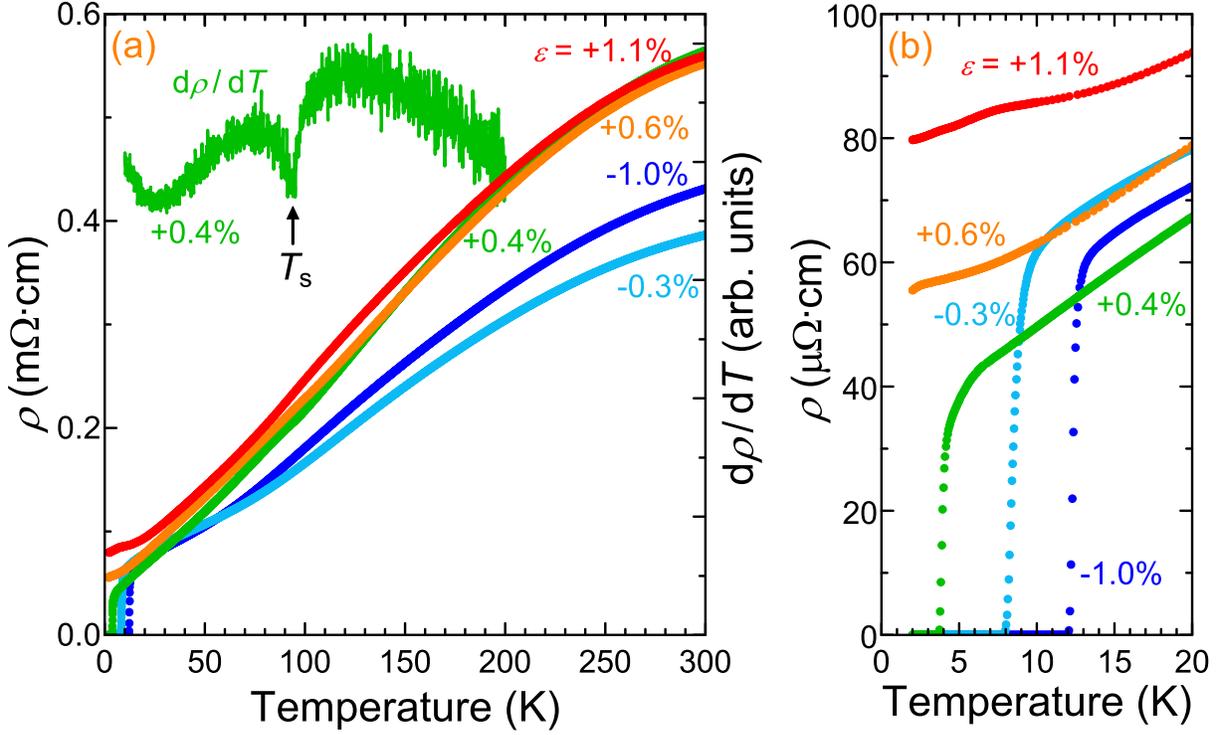}%
\caption{(a) Temperature dependence of the dc electrical resistivity, $\rho$, of FeSe films with various strain. The temperature dependence of  d$\rho /$d$T$ of a film with $\epsilon \sim +0.4$\% is also plotted. (b) Temperature dependence of the resistivity of FeSe films around the superconducting transition temperatures.}
\label{G-RT}
\end{figure}

Figures \ref{G-RT}(a) and (b) show the temperature dependence of the dc electrical resistivity of the grown films.
The residual resistivity ratio, $RRR \equiv \rho(0 \mathrm K) / \rho(300 \mathrm K)$, reached 18 for our films, giving an indication of the good crystalline quality of our films.
Note that a kink anomaly which is due to the structural transition was observed at around 90 K in the $\rho$-$T$ curves, similar to that in bulk samples.
The structural transition temperature, $T_{\mathrm s}$, can be determined from the position of the anomaly in the $\mathrm d \rho / \mathrm d T$ curve (Fig. \ref{G-RT}(a)).
The superconducting transition temperature, $T_{\mathrm c}$, significantly changed depending on the value of $\varepsilon$.
Films with strong compressive strain ($\varepsilon < -1.0$\%) showed $T_{\mathrm c}$ values higher than those of bulk crystals ($T_{\mathrm c}$ = 9 K), while the superconducting transition was not observed for samples with strong tensile strain ($\varepsilon > 0.6$\%).

\begin{figure}
\includegraphics[width=\linewidth, bb=0 0 588 769]{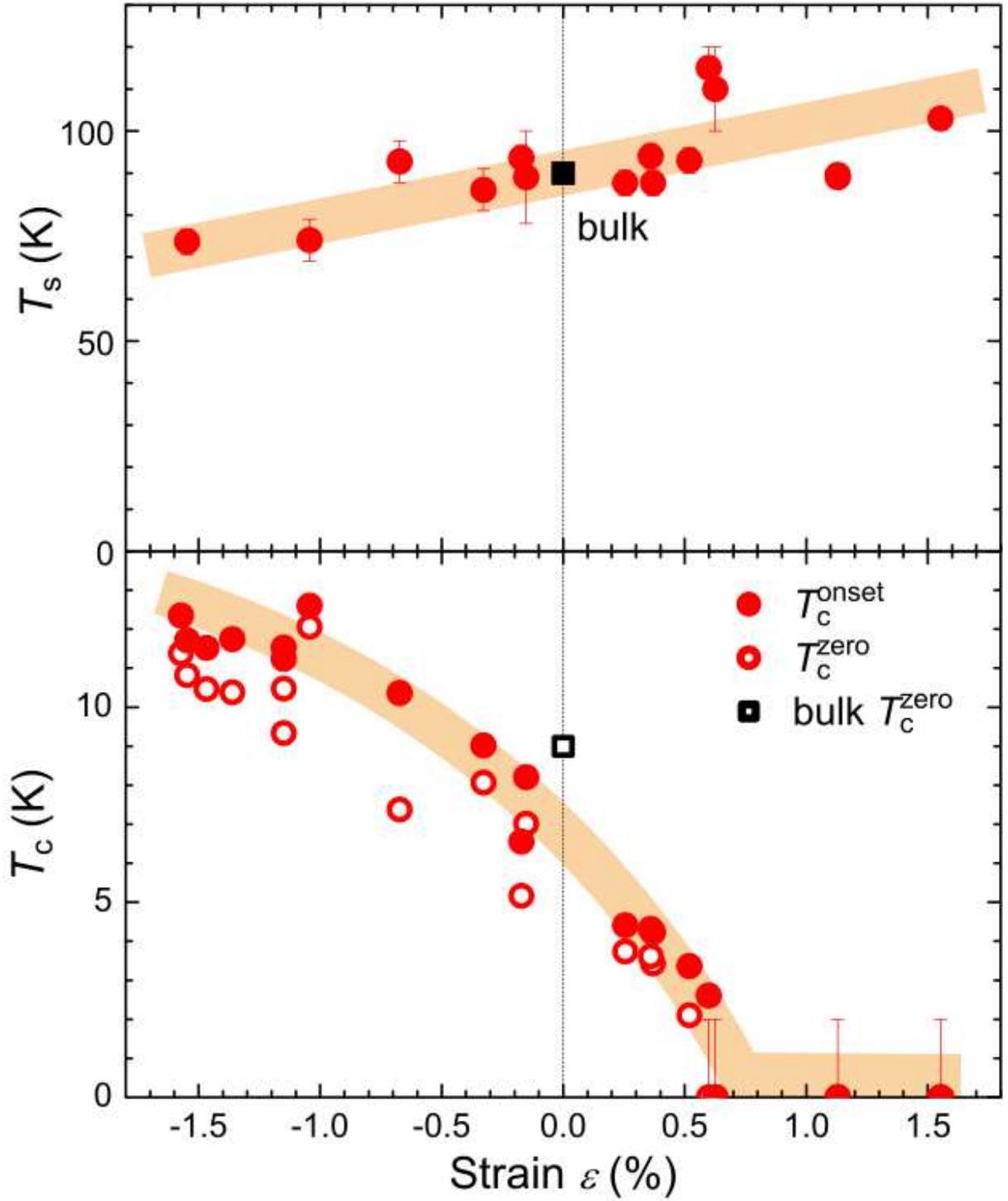}%
\caption{Strain dependence of $T_{\mathrm s}$ (top) and $T_{\mathrm c}$ (bottom) of the FeSe films. Those of bulk samples are also plotted.\cite{PRB.79.014522} The orange lines are guides for the eye.}
\label{G-TsTc}
\end{figure}

Figure \ref{G-TsTc} shows $T_{\mathrm c}$ and $T_{\mathrm s}$ as a function of the in-plane strain parameter, $\varepsilon$.
As already described, when $\varepsilon$ increases from negative to positive, $T_{\mathrm c}$ decreases systematically. 
$T_{\mathrm c}$ drops rapidly in tensile-strained films and films with $\varepsilon > 0.5\%$ do not show zero resistivity above 2 K. 
A recent angle-resolved photoemission spectroscopy (ARPES) study indicates that the rapid decrease in $T_{\mathrm c}$ for films with tensile strain is related to a Lifshitz transition.\cite{Phan17} 
On the other hand, $T_{\mathrm s}$ increases with increasing $\varepsilon$.
In other words, there is a negative correlation between $T_{\mathrm c}$ and $T_{\mathrm s}$ in FeSe under in-plane strain.
This may suggest that the electronic nematicity is unfavorable for raising $T_{\mathrm c}$ in iron chalcogenides, consistent with the sudden increase in $T_{\mathrm c}$ at the disappearance of the nematic transition in Te-substituted films.\cite{yiSciRep17}

The decrease in $T_{\mathrm s}$ due to compressive strain observed for the FeSe films suggests that the structural transition can be extinguished by a smaller amount of Te substitution for films with more compressive strain. %the nematic end point of FeSe$_{1-x}$Te$_x$ will shift to $x=0$ when $\varepsilon$ decreases. 
This is consistent with the previous results in our FeSe$_{1-x}$Te$_x$ films that the $x_{\mathrm c}$ of films on CaF$_2$ is smaller than that of films on LAO, which have longer $a$-axis lengths than films on CaF$_2$. 
As described earlier, considering the fact that (i) the maximum of $T_{\mathrm c}$ is obtained for films with $x \approx x_{\mathrm c}$ and (ii) $T_{\mathrm c}$ increase with decreasing $x$ for $x > x_{\mathrm c}$, it is expected that the realization of smaller $x_{\mathrm c}$ would lead to higher $T_{\mathrm c}$. 
Our results indicate that this is possible by applying more compressive stress in FeSe$_{1-x}$Te$_x$.

\begin{figure}
\includegraphics[width=\linewidth, bb=45 -1 809 592]{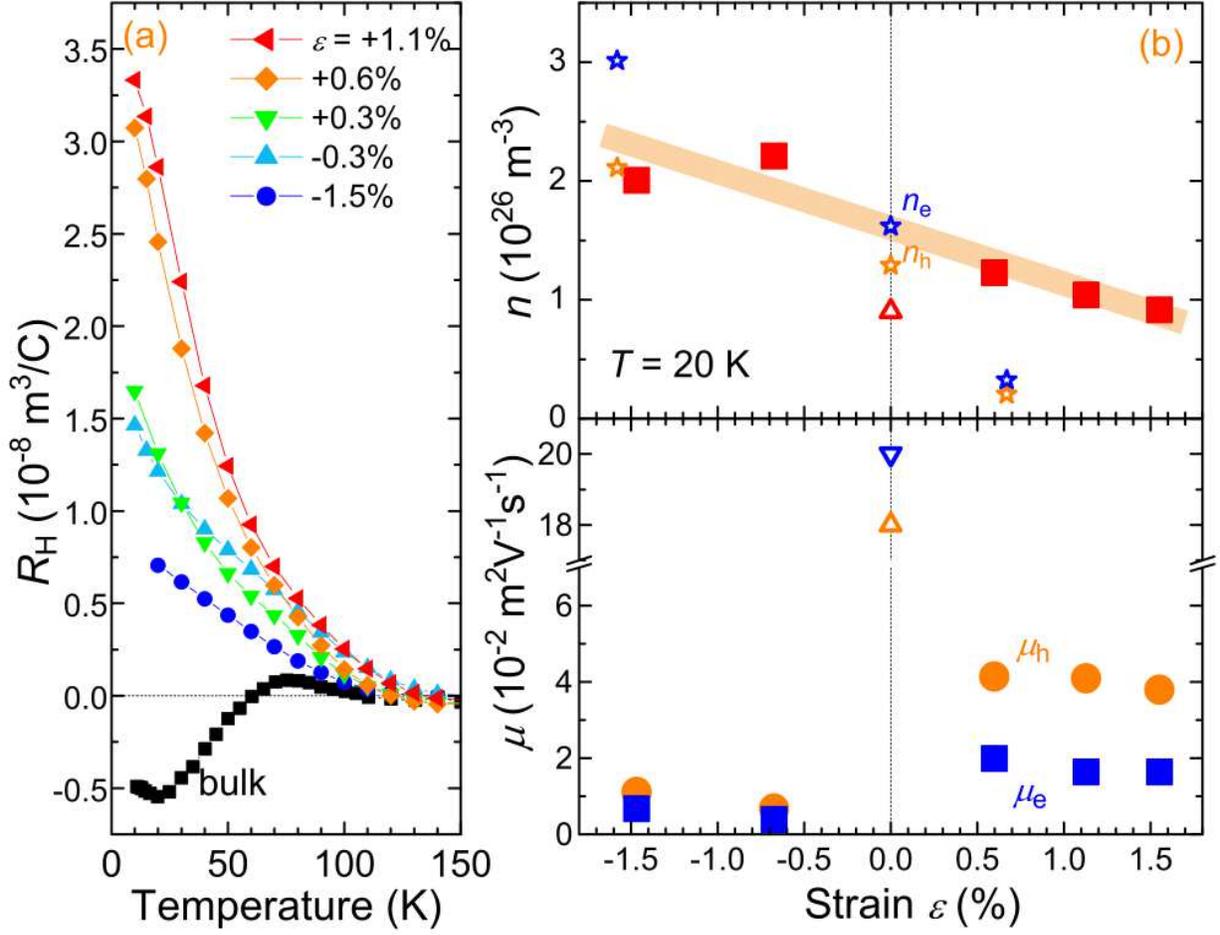}%
\caption{(a) Temperature dependence of the Hall coefficient, $R_{\mathrm H}$, of the FeSe films. Open symbols are data of a bulk sample and  Bulk data\cite{PNAS.111.16309} are also plotted. (b) Strain dependence of the carrier densities (top) and the mobilities (bottom) of the FeSe films. Open triangle simbols represent bulk data,\cite{PhysRevB.90.144516} which were estimated by the same method. Star symbols represents the ARPES results on FeSe bulk and thin film samples.\cite{Phan17}}
\label{G-HallNMu}
\end{figure}

To reveal the nature of charge carriers in the grown films we performed Hall measurements as well as magnetoresistance measurements.
Figure \ref{G-HallNMu}(a) shows the temperature dependence of the Hall coefficient, $R_{\mathrm H}$, of the grown films.
The behavior of $R_{\mathrm H}(T)$ above 100 K is very similar among the samples, including bulk crystals, and the values of $R_{\mathrm H}$ are very small. % among samples????
On the other hand, $R_{\mathrm H}$ increases significantly with decreasing temperature below 100 K, and the strain dependence becomes visible.
$R_{\mathrm H}$ becomes large with increasing $\varepsilon$ at low temperatures. %傾向にある(see fig. %\ref{G_EvsN}(a)
Note that the $R_{\mathrm H}(T)$ of bulk samples deviates from those of films below 100 K, which decreases on cooling below 70--80 K and becomes negative at low temperatures.\cite{PNAS.111.16309}
We will discuss the origins of the difference in the low-$T$ $R_{\mathrm H}$ behavior between films and bulk later.

In a multiband system like iron chalcogenides, where electron- and hole-type carriers coexist, $R_{\mathrm H}$ is not related to the carrier densities in a simple form. 
We considered one electron band and one hole band representing the multiple bands and applied the text-book approach for multiband materials. 
In a classical two-band model, the resistivity tensor is expressed as
\begin{eqnarray}
\rho_{xx} (0) &= \frac{1}{e(n_{\mathrm h}\mu _{\mathrm h}+n_{\mathrm e}\mu _{\mathrm e})},\\
\frac{\rho_{xx} (B) - \rho_{xx} (0)}{\rho_{xx} (0)} &= \frac{n_{\mathrm h}n_{\mathrm e}\mu_{\mathrm h}\mu_{\mathrm e}(\mu_{\mathrm h}+\mu_{\mathrm e})^2}{(n_{\mathrm h}\mu _{\mathrm h}+n_{\mathrm e}\mu _{\mathrm e})^2} B^2, \\
\rho_{yx}(B) &= \frac{n_{\mathrm h}\mu_{\mathrm h}^2-n_{\mathrm e}\mu_{\mathrm e}^2}{e(n_{\mathrm h}\mu _{\mathrm h}+n_{\mathrm e}\mu _{\mathrm e})^2} B,
\end{eqnarray}
where $n_{\mathrm h}$, $n_{\mathrm e}$, $\mu_{\mathrm h}$, and $\mu_{\mathrm e}$ are the hole density, the electron density, the hole mobility, and the electron mobility, respectively.
We evaluated the carrier densities and mobilities of the films with the measured data of $R_{\mathrm H}$ and the magnetoresistance, assuming $n_{\mathrm h} = n_{\mathrm e}$.\footnote{The ARPES measurement revealed that the ratio of the two carrier densities ($n_{\mathrm h} / n_{\mathrm e}$) is $\sim$0.65 - 0.80, and is almost independent on $\varepsilon$.\cite{Phan17} Therefore, even if we use $n_{\mathrm h} / n_{\mathrm e}$ values estimated from the ARPES results when we analyze the magneto-transport data, the obtained results are qualitatively the same.} %In order to evaluate the carrier density and the mobility of the films, we also measured the magneto-resistance. 
%Because there are four parameters for three equations, we assumed the carrier density ratio (nh/ne) based on an angle-resolved photoemission spectroscopy (ARPES) data%[18].

The obtained values of carrier densities and mobilities are plotted as a function of $\varepsilon$ in Fig. \ref{G-HallNMu}(b). 
As $\varepsilon$ decreases, the carrier densities increase. 
This result is consistent with the ARPES results, which also showed the increase in both hole and electron densities for an in-plane compressed sample.\cite{Phan17}
The agreement between our results of the transport measurements and the ARPES study demonstrates that there is a correlation between the $T_{\mathrm c}$ and the carrier densities of FeSe. 
This suggests that the increase in the carrier densities is essential for the increase in $T_{\mathrm c}$. 

On the other hand, we found no significant correlation between both the hole and electron mobilities and $\varepsilon$. 
The fact that the highest $T_{\mathrm c}$ is obtained for films with small mobilities implies that the mobility is not so important for superconductivity in FeSe. 
Note that the hole mobility is always higher than that of electron for our films.
However the opposite behavior was reported for bulk FeSe,\cite{PhysRevB.90.144516} which results in the difference in the sign of $R_{\mathrm H}$ at low temperatures between films and bulk samples.
Although the origin of this difference in mobilities between films and bulk single crystals is unclear at present, we believe that it is not important for superconductivity because the $T_{\mathrm c}$ of the bulk crystal does not significantly deviate from the $T_{\mathrm c}$-vs-$\varepsilon$ curve for our films.
Indeed, a mobility spectrum analysis\cite{PhysRevB.90.144516} revealed that bulk FeSe has minority N-type carriers with very high mobility, while majority of both N- and P-type carriers have comparable mobilities, which may result in higher mobility of electron-like carriers in the two-band model.
These minority carriers with high mobilities are considered to contribute insignificantly to $T_{\mathrm c}$.  

%A mobility spectrum analysis\cite{PhysRevB.90.144516} revealed that in addition to the compensated hole- and electron-type carriers, bulk FeSe has additional minority electron-type carriers with very-high mobilities, which is most likely to be related to Dirac dispersions.
%If somehow film samples do not have such ultrafast electron-type carries,  もしバルクはディラックがあって薄膜になければRhの違いがでてもいいかもしれない
%The  may result in higher mobility of electron-like carriers in the two-band model. 

Finally, we comment on the relationship between the superconductivity and the nematicity in iron chalcogenides.
Our results with strained FeSe suggest that the electronic nematicity is unfavorable for raising $T_{\mathrm c}$ in iron chalcogenides.
As described earlier, this is consistent with the results for the Te-substituted samples, where a rapid increase in $T_{\mathrm c}$ is observed corresponding to the disappearance of the nematic transition.\cite{yiSciRep17}
On the other hand, another isovalent substitution by sulfur also suppresses the nematicity. 
However, there is no significant increase in $T_{\mathrm c}$ when the nematicity disappears, or rather $T_{\mathrm c}$ decreases after the disappearance of the nematicity for S-substituted samples.
The contrasting phase diagrams of FeSe$_{1-x}$Te$_x$ and FeSe$_{1-x}$S$_x$ indicate that the role of the nematicity is not universal in the superconductivity of iron chalcogenides, suggesting that the nematicity affects $T_{\mathrm c}$ only in an indirect manner.\cite{JPSJinpress}
%This difference in the behaviors of $T_{\mathrm c}$ against $T_{\mathrm s}$ among the kinds of applied pressure (anisotropic pressure and positive and negative chemical pressure) indicates that the role of nematicity in superconductivity is not universal in iron chalcogenides.
Rather, our results suggest that the most essential factor for realizing high $T_{\mathrm c}$ is the increase in the carrier densities.

This conclusion may be inconsistent with our previous results of Hall measurements with FeSe$_{1-x}$Te$_x$ films,\cite{Tsukada10,it15jjap,ys16jpsj} which suggested that the $n_{\mathrm h}$-to-$n_{\mathrm e}$ and/or $\mu_{\mathrm h}$-to-$\mu_{\mathrm e}$ ratio were essential for high $T_{\mathrm c}$.
This disagreement may suggest that there are multiple channels for increasing $T_{\mathrm c}$, originating from the multiband/multiorbital character in iron chalcogenides.\cite{NCommun.8.14988}
In other words, it is necessary to increase carrier densities for obtaining high $T_{\mathrm c}$ of up to 12 K in FeSe, and for further enhancement of $T_{\mathrm c}$ of up to 23 K in FeSe$_{1-x}$Te$_x$ we may need to tune the ratio of carrier densities and/or mobilities. %of we may need to tune the carrier densities for enhancing Tc up to $\sim$12 K and  
Further comprehensive and systematic studies with Te- and S-substituted samples are needed for a complete understanding of the behavior of $T_{\mathrm c}$ in iron chalcogenides, which is now under way.

In conclusion, we have succeeded in growing a series of FeSe films with various degrees of the in-plane strain, from tensile to compressive.
We found that as the strain becomes more compressive, the structural transition temperature decreases.
This result suggests that the difference of the substitution content $x$ that is required for the complete suppression of the structural transition between FeSe$_{1-x}$Te$_x$ films on LAO and on CaF$_2$ is due to the difference in the degree of strain.
This means that the structural transition can be extinguished by a smaller amount of Te substitution for films with more compressed strain, which may lead to higher $T_{\mathrm c}$. 
It was also found that $T_{\mathrm c}$ and the carrier densities of the FeSe films increase systematically as $\varepsilon$ decreases. 
The clear correlation between $T_{\mathrm c}$ and the carrier densities suggests that for the $T_{\mathrm c}$ enhancement of iron chalcogenides it is essential  to increase the carrier densities.

\begin{acknowledgments}
We would like to thank K. Ueno at the University of Tokyo for the X-ray measurements.
We also thank M. Hanawa at Central Research Institute of Electric Power Industry for his support in the thickness measurement.
Finally, we are grateful to Editage (www.editage.jp) for English language editing.
\end{acknowledgments}

\clearpage

\clearpage

\end{document}